**Using hierarchical statistical learning models to model individual statistical learning**

Hanna Ringer, Tatsuya Daikoku*

*Next Generation Artificial Intelligence Research Center, Graduate School of Information Science and Technology, The University of Tokyo, Tokyo, Japan*

** Corresponding author: Tatsuya Daikoku, daikoku.tatsuya@mail.u-tokyo.ac.jp*

**Abstract**

Statistical learning is essential for individuals to discover structure in the sensory environment, especially during communication via speech or music. Individual differences in statistical learning abilities have been proposed to account for differences in various cognitive functions and development, including developmental disorders such as dyslexia. In this study, we used a Hierarchical Bayesian Statistical Learning (HBSL) model to model individual learning trajectories as recorded using electroencephalography (EEG) while adults with and without dyslexia listened to structured tone sequences. Although we did not find a significant group difference, our results showed a close correspondence of between the model simulations and the real EEG data and novel sequences generated based on individual models were highly similar to the original stimulus sequence. This provides a proof of concept for future research and suggests that the HBSL model accurately represented the statistical sequence structure in a similar way as did human listeners.

**Acknowledgements**

This research was supported by the Japan Society for the Promotion of Science (JSPS), KAKENHI (25K03206, 25K24741, 26K23870), National Institute for Japanese Language and Linguistics (https://dx.doi.org/10.13039/100018205), a Postdoctoral Fellowship for Research in Japan (PE24035) by the JSPS, and an AI Fusion Research Promotion Fund by the Next Generation Artificial Intelligence Research Center at The University of Tokyo awarded to Hanna Ringer. The funding sources had no role in the design of this study, data analysis and interpretation, and the decision to publish this manuscript.

## Introduction

Statistical learning is a powerful innate mechanism to acquire knowledge about the statistical structure of the environment (Saffran, 2003; Saffran et al., 1996). In particular, in sequential signals such as language or music, statistical learning enables the segmentation of the signal into meaningful units, e.g., the segmentation of a continuous speech stream into single words or a musical melody into distinct phrases, based on transitional probabilities between single elements within the sequence, e.g., syllables or tones (Daikoku, 2018). Thus, efficient statistical learning is essential for successful language acquisition and communication. In fact, inter-individual differences in statistical learning abilities have been associated with various cognitive functions and development. For example, deficits in statistical learning were reported in individuals with developmental disorders such as autism spectrum disorder (ASD), which is often linked to communication difficulties, attention-deficit/hyperactivity disorder (ADHD), and developmental dyslexia (Daikoku, Kamermans, et al., 2023; Obeid et al., 2016; Saffran, 2018).

The predictive coding framework (Clark, 2013; Friston, 2005; Friston et al., 2017) postulates continuous anticipation of sensory input, with the aim to minimise both the so-called prediction error, i.e., the mismatch between the predicted and the actual input, and perceptual uncertainty, i.e., increase the reliability of the internal predictions. Based on this framework, simulation studies attempted to explain individual differences in cognitive development and functioning as a result of differences in how individuals weigh their prior assumptions and sensory input while making predictions about upcoming events (e.g., Pellicano & Burr, 2012; Philippsen et al., 2022). Individuals' focus on prior assumptions and their sensitivity to recent input shape how information is perceived, represented, and how new information is generated: Individuals who put a stronger weight on priors and are less sensitive to sensory information tend to base their interpretation of the current input more strongly on previous experiences, whereas individuals who put less weight on priors and are more sensitive to sensory information quickly pick up changes in dynamic environments. These individual differences in weighting of prior information and sensitivity to new input were in turn related to various developmental disorders, including ASD (e.g., Pellicano & Burr, 2012; Van de Cruys et al., 2014) and ADHD (e.g., Bijlenga et al., 2017; Panagiotidi et al., 2018), as well as affective disorders (e.g., Engel-Yeger et al., 2016). Statistical learning may constitute an important link between an individual's sensory processing of statistical regularities, i.e., their sensitivity to sensory statistics and tendency to form meaningful chunks in continuous sequential signals, and higher-level cognitive functions and development.

As mentioned above, a growing body of research has provided evidence that statistical learning is impaired in individuals with dyslexia, which is a specific learning disorder characterised by deficits in reading and spelling acquisition, even though individuals have normal intelligence and receive adequate school education (Lyon et al., 2003; Peterson & Pennington, 2015). The reported deficits in processing structure in the (auditory) environment range from reduced adaptation to simple repetitions of voices or tones (Perrachione et al., 2016; Peter et al., 2019) to impaired extraction of more complex statistical regularities such as transitional probabilities (Gabay et al., 2015) or distributional statistics (Kimel et al., 2022). In particular, several studies demonstrated that neural tracking of the statistical structure in continuous auditory sequences is diminished in individuals with dyslexia. These studies used an established statistical learning paradigm in which participants are passively presented with a constant stream of syllables that consisted of four tri-syllabic artificial "words", which were repeated in random order (Saffran et al., 1996).

Transitional probabilities between adjacent syllables, which were higher within words and lower between words, were the only cue to segment the continuous stream and chunk single elements into meaningful units, i.e., boundaries between distinct words. In typically developing individuals, successful extraction of the statistical structure is reflected in an increased synchronisation of neural oscillations, i.e., inter-trial coherence (ITC) at the word rate relative to the single-syllable rate as measured using electroencephalography (EEG; Batterink & Paller, 2017, 2019). Individuals with dyslexia, by contrast, showed relatively weaker neural tracking of the word structure (Ringer et al., 2025; Zhang et al., 2021) and a reduced mismatch response to unexpected deviations from the statistical structure, i.e., when the last syllable of a word was unexpectedly replaced by a different syllable (Daikoku, Jentschke, et al., 2023). Notably, reduced neural tracking of the statistical structure was also found when tones with different frequencies that were arranged into triplets were used instead of syllables, suggesting that individuals with dyslexia have a broader deficit in processing structure in sound sequences that is not specific to linguistic material (Ringer et al., 2025).

While information about behavioural and neural differences between groups of participants, e.g., individuals with and without a formal diagnosis of a developmental disorder, provides valuable insight into the association of statistical learning with development and cognitive functions, it is also crucial to look into individual learning trajectories to gain a deeper understanding of the mechanisms underlying these links. Computational modelling offers a promising approach to model these individual learning trajectories and identify parameters that explain statistical learning differences between individuals or groups of individuals, which may in turn inform about the underlying mechanisms. In the present study, we used computational modelling to model neural tracking of the statistical structure of a tone sequence as measured using EEG in a previous study in individuals with and without dyslexia (Ringer et al., 2025). We used the modelling results to examine differences in terms of sensitivity to the sensory input and the tendency to form chunks of adjacent tones based on transitional probabilities as well as to probabilistically generate new sequences based on individual models. In accordance with previous studies that reported impaired statistical learning in dyslexia, it is plausible to assume that individuals with dyslexia show a lower sensitivity to the sensory input and/or a higher chunk threshold, i.e., a lower tendency to chunk adjacent elements into one unit, both of which result in slower learning of the statistical sequence structure. Moreover, we hypothesised that sequences that were produced based on models from the control group contained more chunks than those produced from the dyslexia group, which reflects a stronger representation of the statistical triplet structure.

## Methods

In this study, we combined EEG data from a previous study (Ringer et al., 2025) with computational modelling to model individual learning, compare model parameters between individuals, and probabilistically generate sound sequences based on individual models. We used a Hierarchical Bayesian Statistical Learning (HBSL) model (Daikoku, Kamermans, et al., 2023), which integrates Bayesian estimation with a Markov model, using a Dirichlet distribution as a prior distribution. This model incorporates the reliability of the transitional probabilities between elements within a sequence, calculated from the inverse of the variance of the prior distribution of the transitional probabilities. Transition patterns are chunked into one unit if the product of normalised probability and reliability exceeds a certain chunk constant, i.e., when both the transitional probability and the reliability of the probability estimate are high.

We manipulated two parameters in our model simulations to account for inter-individual differences in statistical learning: First, we varied the degree of dependence on the sensory signals (relative to prior distributions) by multiplying the parameter vector of the Dirichlet distribution with a sensitivity constant. Hypo-sensitive models (sensitivity constant < 1) relied more on prior predictions, whereas hyper-sensitive models (sensitivity constant > 1) relied more on the sensory input and therefore learnt the statistical structure more quickly. Second, we varied the threshold that the product of probability and reliability has to reach for the model to chunk two adjacent elements into one unit. Models with a low chunk constant had a high tendency to chunk and quickly segmented the tone sequence according to the underlying triplet structure, whereas models with a high chunk constant had a lower tendency to chunk.

Our methodological procedure is illustrated in Figure 1. We generated a set of simulated EEG datasets based on the model results obtained with different parameter combinations. For both sensitivity and chunk constants, a set of eleven values was defined, which should cover a biologically plausible range, resulting in a total of 121 possible parameter combinations. For any of the combinations, the stimulus sequence used in the EEG study (see Figure 1 in Ringer et al., 2025) was submitted to the HBSL model. For each element in the sequence, the resulting model output provided the transitional probability, its reliability, and a binary variable reflecting whether the model formed a chunk with the preceding element. An EEG waveform was simulated by adding random white noise, a sinusoidal oscillation at the tone rate (3.33 Hz), and oscillations at the rate of tone triplets (1.11 Hz) whenever the model chunked these elements. From the simulated EEG data, we computed the time-resolved triplet learning index (TLI), reflecting the strength of perceptually tracking triplets relative to single tones, in the same way as described for the real data in Ringer et al., 2025. We correlated each individual's real EEG data with the 121 simulated TLI time courses to identify the best-fitting model (with the highest correlation) and extract its sensitivity and chunk constants. To avoid overfitting to random noise in the simulated EEG data, we repeated the simulation 1000 times and averaged the extracted parameters across the 1000 runs. The averaged parameters were then compared between individuals with ($n$ = 17) and without ($n$ = 18) dyslexia, using independent $t$-tests. Finally, we probabilistically generated novel tone sequences based on the transitional probability distributions obtained from each individual's best-fitting model, using an automatic composition process (Daikoku & Nagai, 2022). Probability distributions were averaged across 30 generated sequences per individual and compared with the original (presented) sequence using the Kullback-Leibler divergence (which was used in previous studies to measure the similarity of probability distributions, e.g., Daikoku, 2023). Moreover, we calculated the number of chunks in the generated sequences, averaged it across the 30 sequences, and compared it between the control and the dyslexia group.

## 1 Model-based simulation of EEG data

For each combination of sensitivity and chunk constants ($n$ = 121):

**Step 1:** submit stimulus sequence to HSL model

A B C D E F A B C ...

**Step 2:** simulate EEG data based on model output

**noise**

+

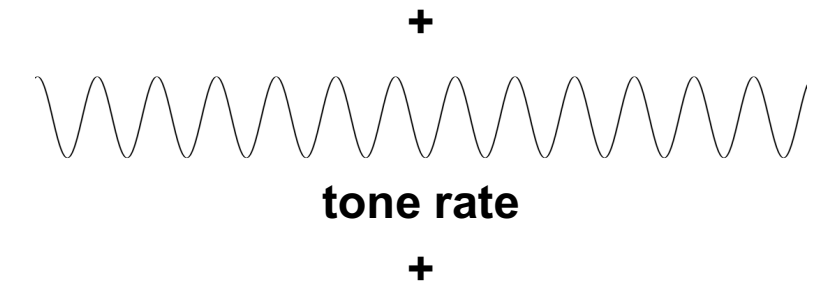

**tone rate**

+

**triplet rate**
(whenever the model chunks three adjacent tones)

**Step 3:** compute TLI over time from simulated data

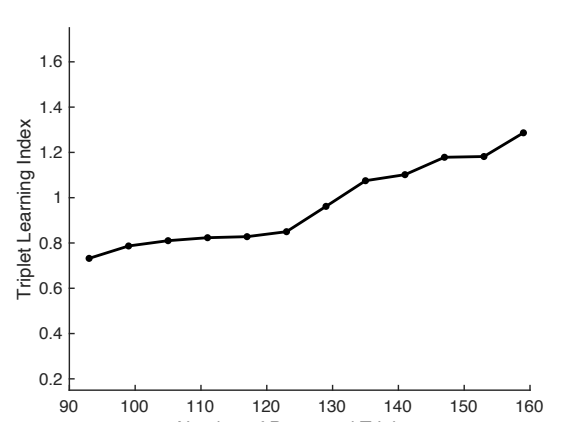

## 2 Comparison of simulated and real EEG data

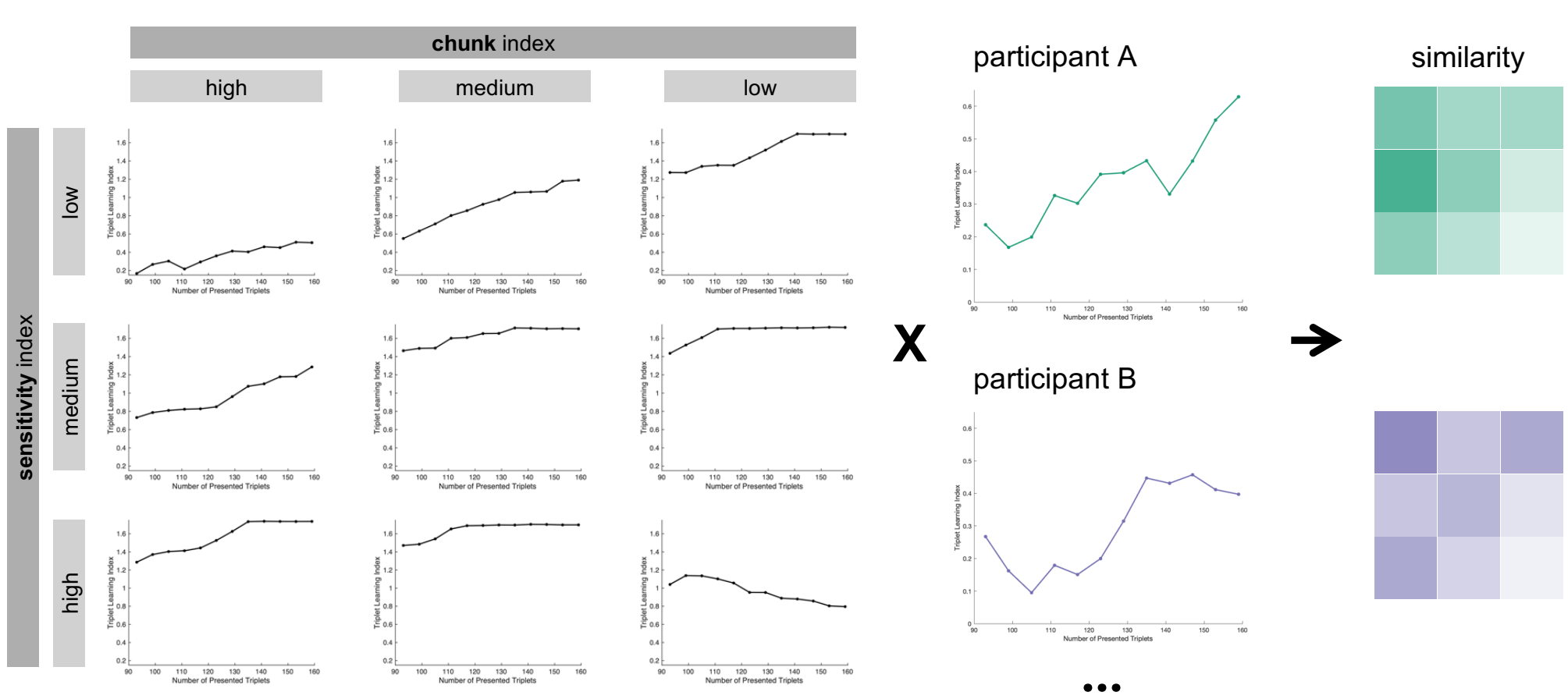

## 3 Group comparison of model parameters

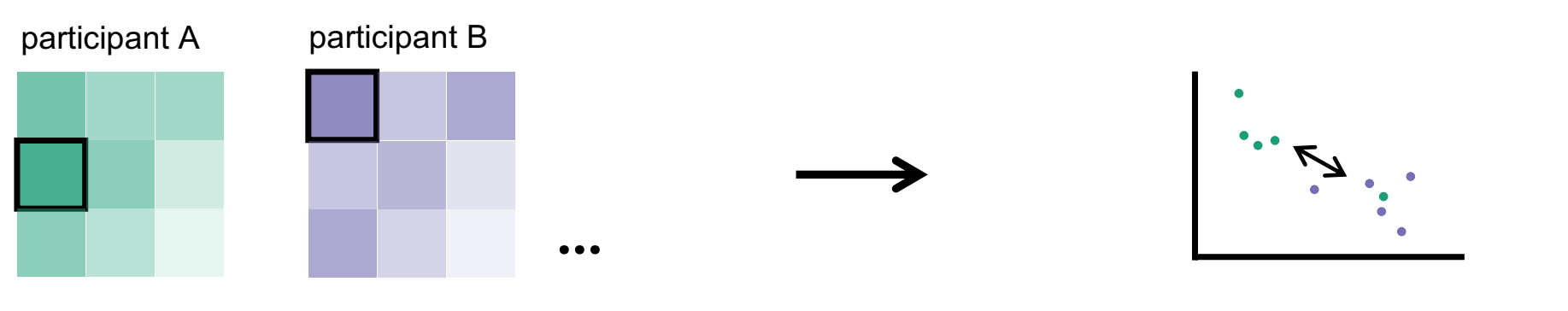

extract sensitivity and chunk constants from the best-fitting model for each participant ($n$ = 35)

compare model parameters between control and dyslexia group

## 4 Probabilistic generation of novel sequences

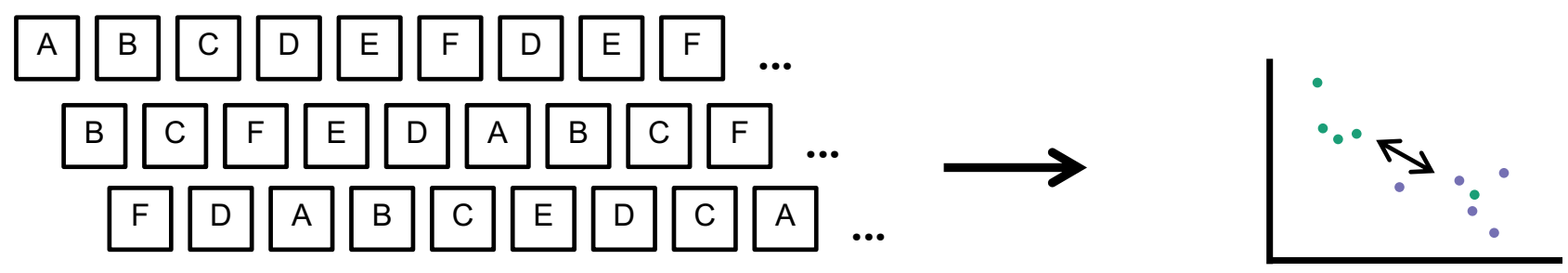

generate novel sequences using an automatic composition process based on individual probability distributions

compare number of chunks in generated sequences between control and dyslexia group

*Figure 1.* Illustration of the methodological approach. Please note that the nine exemplary simulated triplet learning index (TLI) time courses and the real TLI time courses from two exemplary participants (panel 2) only serve illustrative purposes. In the actual analysis, we used a larger number of simulated TLI time courses ($n$ = 121) and compared them with each participant's ($n$ = 35) real data.

## Results

Results are shown in Figure 2. The mean maximum correlation between simulated and real EEG data (see Figure 2A), i.e., the correlation of the TLI time course in the real data with the one of the best-fitting model (averaged across all 1000 EEG data simulation runs), was 0.89 $\pm$ 0.14 ($M \pm SD$). The correlation was slightly lower for the dyslexia group (0.84 $\pm$ 0.19) than for the control group (0.93 $\pm$ 0.06); however, an independent $t$-test did not reveal a significant group difference ($t(18) = 1.72$, $p = .102$, $d = 0.59$). Across all participants, the two parameters extracted from the best-fitting model were negatively correlated ($r = -.68$, $t(34) = -5.37$, $p < .001$), such that individuals with a higher sensitivity to the statistical structure of the sequence had a lower chunk threshold, i.e., a higher tendency to chunk adjacent tones into triplets (see Figure 2B).

When comparing the parameters of the best-fitting models between groups (see Figure 2C and 2D), the control group (1.15 $\pm$ 0.36) tended to show higher sensitivity than the dyslexia group (1.06 $\pm$ 0.39). However, this group difference fell short of statistical significance ($t(32) = 0.70$, $p = .246$, $d = 0.23$). With regard to the tendency to chunk adjacent elements within the sequence, the control group (10.49 $\pm$ 6.77) showed a lower chunk threshold, i.e. a higher tendency to chunk, than the dyslexia group (12.97 $\pm$ 7.44). Again, as for the sensitivity parameter, the group difference did not reach statistical significance ($t(32) = -1.04$, $p = .153$, $d = 0.35$).

## Parameter distribution

**A Correlation of real & simulated data**

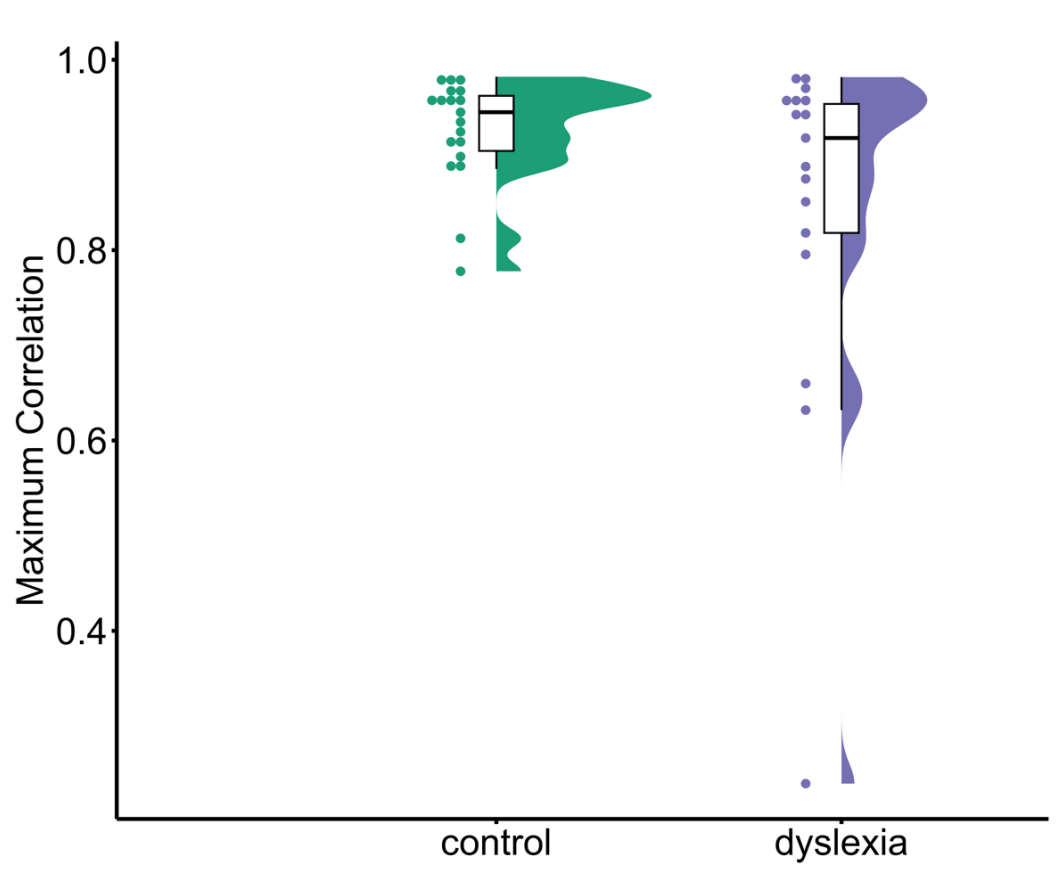

**B Parameter correlation**

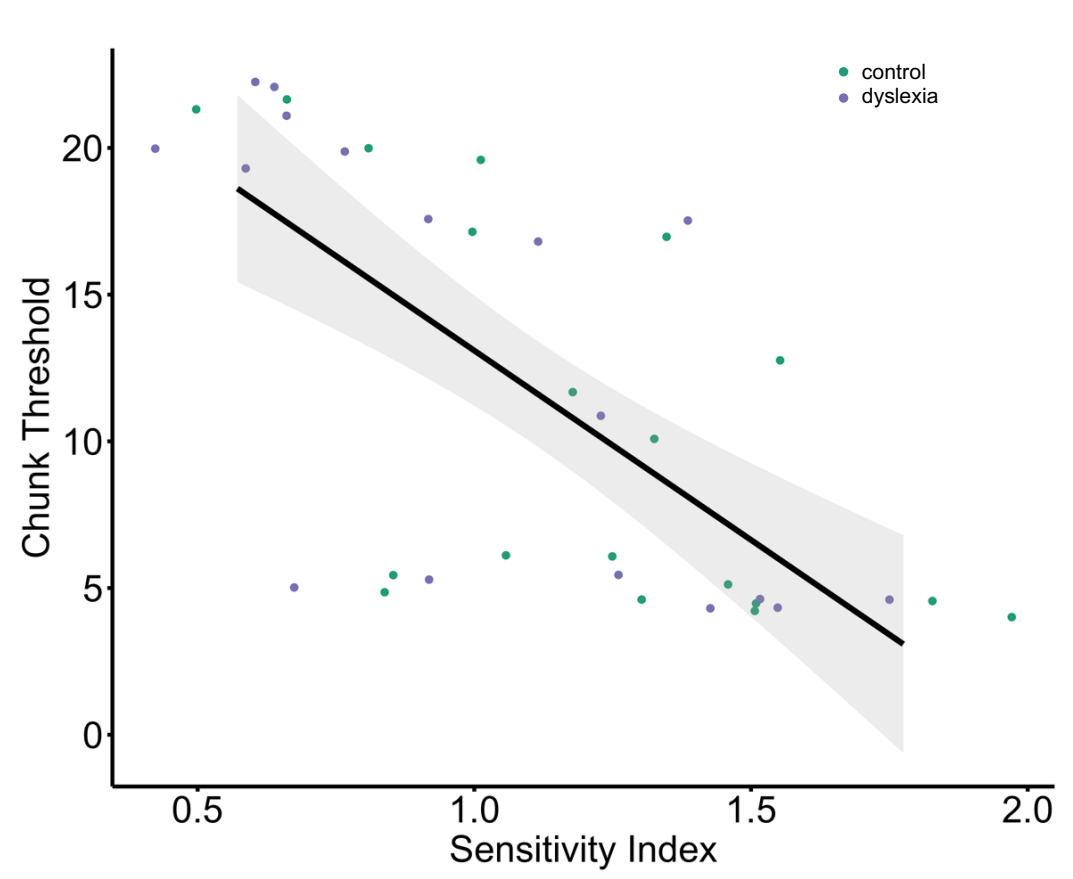

## Statistical sequence learning

**C Sensitivity constant**

Sensitivity Index
1.8
1.5
1.2
0.9
0.6
control
dyslexia

**D Chunk constant**

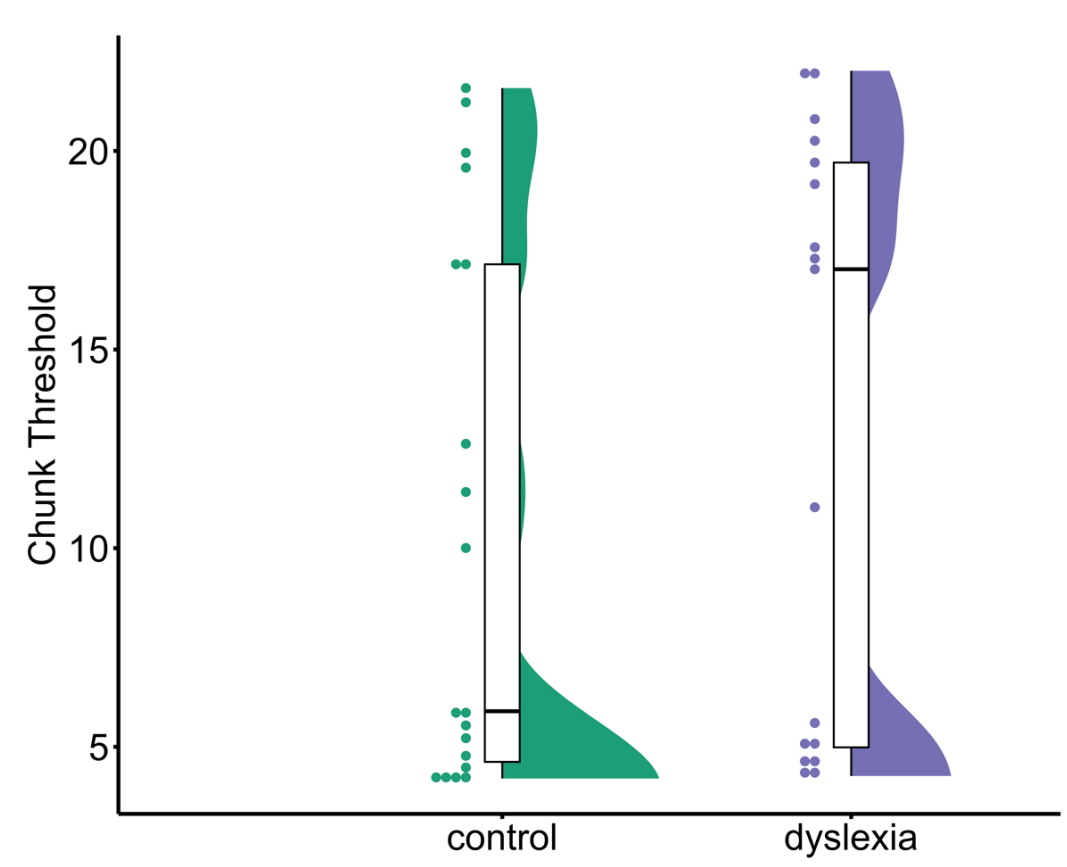

## Probabilistic sequence generation

**E Kullback-Leibler divergence**

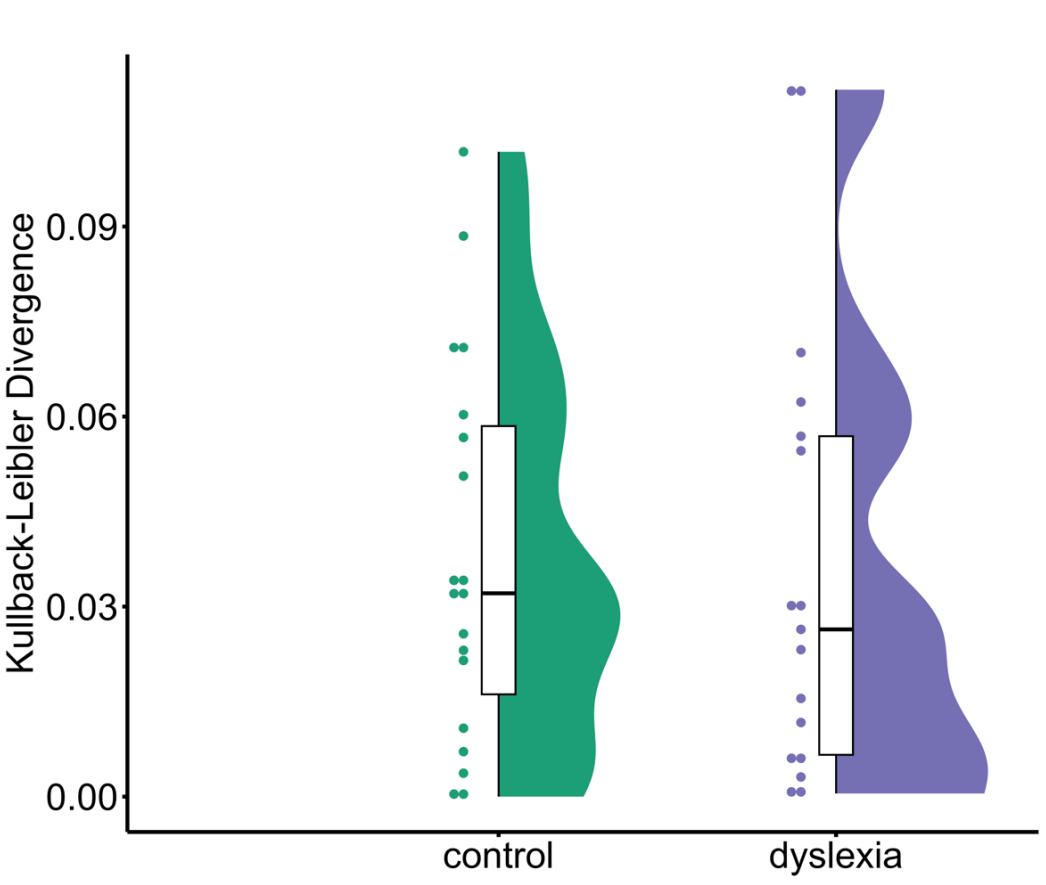

**F Number of chunks**

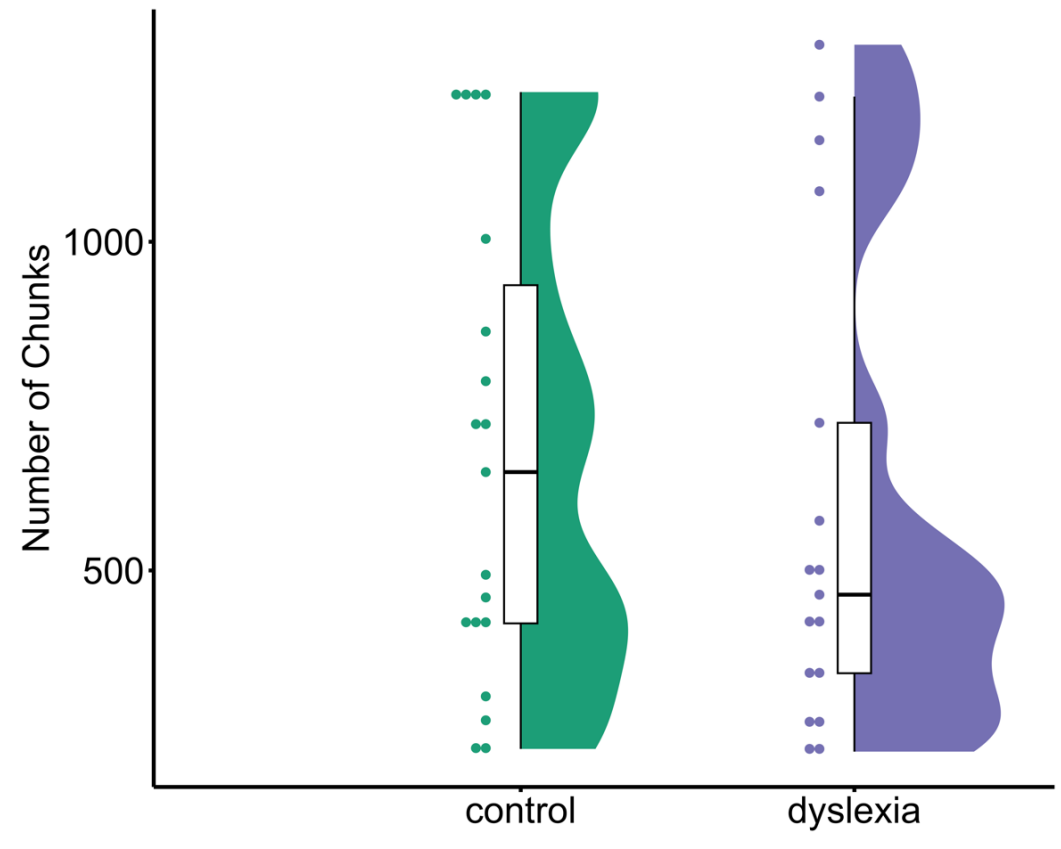

*Figure 2.* A: Correlation of real and simulated EEG data from the best-fitting model (averaged across all 1000 simulation runs). B: Correlation between Sensitivity Index (x-axis) and Chunk Threshold (y-axis) as extracted from the best-fitting model for each participant. Shaded areas around the trend line indicate the 95 % confidence interval. C & D: Sensitivity Index (panel C) and Chunk Threshold (panel D) as extracted from the best-fitting model for each participant. E: Kullback-Leibler Divergence as a measure of difference between the original sequence and the probabilistically generated novel sequences (averaged across all 30 generated sequences). Lower values close to zero indicate highly similar probability distributions. F: Number of chunks contained in the probabilistically generated novel sequences. Across all plots, single data points correspond to individual participants, and boxes correspond to the first quartile (lower border of the box), median (bold line inside the box), and third quartile (upper border of the box), respectively.

The novel sequences generated using an automatic composition process based on individual probability distributions were overall highly similar to the original sequence, as indicated by a low Kullback-Leibler Divergence with values very close to zero ($0.04 \pm 0.03$), with no significant difference between groups ($t(32) = 0.15$, $p = .883$, $d = 0.05$). The probabilistically generated sequences tended to contain more chunks for the control ($679 \pm 360$) than for the dyslexia group ($591 \pm 367$). However, the group comparison did not reveal a statistically significant difference ($t(32) = 0.73$, $p = .237$, $d = 0.24$).

## General Discussion

This study aimed to explore inter-individual differences in auditory statistical learning by combining EEG data from a previous study (Ringer et al., 2025) with computational modelling in a sample of participants with and without a diagnosis of developmental dyslexia. The results showed an overall high correlation of the simulated and the real EEG data across all participants, and an automatic composition algorithm produced novel sequences that were highly similar to the original sequence based on probability distributions of the individual models. However, even though group differences between individuals with and without dyslexia pointed in the expected directions for both model parameters, i.e., the sensitivity to the statistical structure and the threshold to chunk adjacent elements into meaningful units, and the complexity of the probabilistically generated novel sequences, i.e., the number of chunks contained in those sequences, statistical comparisons did not yield significant effects.

The high correlation between the real and the simulated data provides preliminary support that the HBSL model is suited for modelling human statistical learning as reflected in neural synchronisation with the statistical stimulus structure. Moreover, the high similarity between the probabilistically generated novel sequences and the original stimulus sequence implies that the model adequately represents the statistical properties of the stimulus input. The strong correlation between the two parameters suggests that logically plausible models were mostly selected as an individual's best-fitting model, e.g., models with a higher sensitivity to the statistical structure and a lower chunk threshold, i.e., a stronger tendency to chunk single tones into triplets, or models with a lower sensitivity and a higher chunk threshold. Conversely, implausible models, e.g., with a high sensitivity and high chunk constant, were only rarely selected as the best-fitting model. Together, these insights highlight the potential of using the HBSL model for modelling individual statistical learning and explain inter-individual differences in behavioural and neural correlates of learning in future research.

Even though the comparisons between groups did not yield statistically significant effects, the descriptive differences between the dyslexia and the control group were in line with our hypotheses. Specifically, individuals with dyslexia tended to show a lower sensitivity to the statistical structure of the sequence and less chunking of single tones into triplets. Moreover, the novel sequences that were generated based on the probability distributions of their individual best-fitting models tended to contain fewer chunks compared with individuals without dyslexia. Together, these results may point towards impaired processing of (auditory) statistical regularities, which is in line with our previous electrophysiological findings of weaker neural tracking of the statistical sequence structure (Ringer et al., 2025) and diminished responses to unexpected violations of the statistical rules (Daikoku et al., 2023) in individuals with dyslexia. Along with a growing body of behavioural and neural evidence (e.g., Gabay et al., 2015; Perrachione et al., 2016; Zhang et al., 2021), they provide empirical support for various theories that have associated phonological deficits in dyslexia with broader deficits in statistical learning (e.g., Ahissar, 2007; Banai & Ahissar, 2018; Goswami, 2011, 2019; Singh & Conway, 2021; Szmalec et al., 2011).

The absence of a statistically significant group difference between individuals with and without dyslexia may be grounded in the stimulus design. The stimulus sequences used in our study were rather simple and consisted of only two tone triplets, i.e., six different tones in total. It is plausible to assume that individuals with dyslexia were able to extract this simple statistical structure after sufficient exposure to the stimulation (to a similar extent as did individuals without dyslexia), but would have more difficulty in processing more complex statistical regularities. In fact, there is evidence that individuals with dyslexia in particular struggle to process more complex, temporally irregular rhythms, while they are able to extract the beat from simpler, temporally regular rhythms to the same extent as the control group (Fiveash et al., 2020). A similar pattern may hold true for statistical regularities, such that group differences in HBSL model parameters would become more apparent for more complex statistical regularities, e.g., including a larger number of different elements or hierarchical rules. Moreover, the relatively fast stimulation in the EEG experiment resulted in an overlap of event-related responses to the single tones, which may have led to a considerable amount of noise in the data at the single-subject single-trial level (which was used in the present analysis) and could account for some random variability within groups. To gain further insight into the computational bases of individual differences in statistical learning abilities, future studies may vary the complexity of the statistical regularities in the stimulus material. It is plausible to assume that individual differences become particularly obvious when approaching the limits of human processing capacity, e.g., by using a large number of individual elements, a larger number of chunks, chunks of varying length, or hierarchical relationships between chunks. Additionally, implementing statistical regularities that resemble regularities in natural stimulus material such as a speech or music, e.g., as extracted from a speech or music corpus, may allow to further explore the role of statistical learning for processing communication-relevant auditory signals. This also offers the opportunity to study cross-cultural effects of socialisation with speech and music that contain distinct statistical properties. Finally, it may be interesting to include further individual characteristics, such as working memory capacity, ideally as continuous variables, that may account for inter-individual variability in statistical learning besides a binary diagnosis of dyslexia based on phonological skills.

In conclusion, our study provides a proof of concept on how the HBSL model can be used in future research to investigate the computational bases underlying inter-individual variability in statistical learning – a skill that is crucial for efficient perception, cognition, and communication.